\documentclass[onecolumn,noshowpacs,nofootinbib,colorlinks,hyperindex,unicode,11pt]{revtex4}
\usepackage{graphicx}
\usepackage{amsfonts,amsmath,amssymb,tikz,accents}
\usepackage[eulergreek]{sansmath}
\usepackage{indentfirst,bigints}
\usepackage{bm,graphics,color}
\usepackage{feynmp-auto}
\usepackage{slashed}
\counterwithout{equation}{section}
\usepackage{pgfplots}
\pgfplotsset{compat=1.16}

\usepackage{textgreek}
\usepackage{color,xcolor}

\def\0{\mbox{\boldmath$\displaystyle\mathbb{O}$}}

\def\p{\partial}

\usepackage{float,xcolor,upgreek}
\usepackage{color} 
\newcommand{\beq}{\begin{eqnarray}}
\newcommand{\eeq}{\end{eqnarray}}
\newcommand{\bea}{\begin{eqnarray}}
\newcommand{\eea}{\end{eqnarray}}

\begin{document}

\title{Covariant operator formalism for higher derivative systems: \\
Vector spin-$0$ dual model as a prelude to generalized QED$_4$}

\author{G. B. de Gracia}
\email{gabriel.gracia@uftm.edu.br}
\affiliation{Physics Department, Federal University of the Triângulo Mineiro UFTM,\\
38.025-180, Uberaba, MG, Brazil}

\author{A. A. Nogueira}
\email{anderson.nogueira@unifal-mg.edu.br}
\affiliation{Institute of Exact Sciences, Federal University of Alfenas UNIFAL,\\   Alfenas 37133-840, Brazil}


\begin{abstract}
\noindent
In this work we extend the Kugo-Ojima-Nakanishi covariant operator formalism to quantize two higher derivative systems, considering their extended phase space structures. More specifically, the one describing spin-$0$ particles by a vector field and the generalized electrodynamics. We investigate the commutator structure of these theories and present the definition of their physical Hilbert subspaces. The
first model presents a reducible gauge symmetry, implying the necessity of two sets of auxiliary fields. The massless limit is also carefully analyzed. After this prelude, the generalized QED$_4$ can be investigated with such machinery. Regarding the interacting regime, the positive norm subspace is no longer time invariant, since the interaction can create negative norm states from an initially ghost-free one. Then, we furnish an alternative description of the situation by analyzing a set of spectral representations highlighting the lack of positivity associated with the well-known ultraviolet improvement.  Finally, based on these efforts and also on recent discussions about Lee-Wick like models, we prove that it is possible to establish a specific higher derivative interacting model compatible with establishing a time-invariant positive norm subspace.
\end{abstract}

\maketitle


\section{Introduction}

\indent Higher derivative field theories can hardly be overestimated if one considers the wide range of their applications in different areas of physics. Regarding field theories, the inclusion of higher derivative operators can also be understood as quantum corrections \cite{barw}. Accordingly, as in \cite{PV1,PV2,PV3,PV4,LW6,LW7,LW8,tesw}, the  Lee-Wick model, a higher derivative field theory, can be regarded as a system obtained by elevating the status of the Pauli-Villars regulators to dynamical degrees of freedom. Another pertinent model correlated to this context is the so-called Bopp-Podolsky Lagrangian \cite{pode1,pode2,pode3,pode4,pode5,pode6}. It leads to a generalized electrodynamics whose associated static potential is well-behaved at the origin \cite{LW9,LW10,aaw}. Moreover, as an ultraviolet improvement \cite{pod1w,pod2w,pod3w,pod4w,pod5w,pod6w,pod7w, and, pod8w}, it also leads to a finite self-energy for the electron.\\
\indent These kinds of features based on a higher derivative Lagrangian can occur in a set of alternative contexts such as in one loop renormalized  linearized quantum gravity, for example. The generation of higher derivative quantum corrections can be directly verified from the functional renormalization group analysis \cite{cod1w,cod2w,cod3w}. One can also cite the alternative framework of the string theory in which a generalized gravitational theory including higher derivative terms, associated with the product of curvatures, emerges in the low energy limit \cite{strw,strw1}. Although these aforementioned models may be associated with the presence of extra ghost particles with negative norms \cite{shap,quad}, there are different manners to deal with such Ostrogadskian instabilities \cite{grav2,grav}. We also mention the importance of the Lee-Wick models in gravity \cite{g1,g2,g3,g4,g5,g6,g7,f2,f3,f4} due to the mechanisms that conciliate renormalizability with unitarity. Finally, it is worth mentioning the rich and longevous discussion on these ghosts \cite{G1,G2,G3,G4,G5,G6,G7,G8,G9,G10,G11}, in which its general properties are investigated. \\
\indent The other fundamental issue of the paper is closely related to duality in field theories. Namely, using the master action approach one can derive dual relations between theories describing spinless particles using a scalar field and the ones associated with a second derivative order vector field description \cite{d1w,d2w}. It is worth mentioning that the last formulation presents no particle content at the massless limit. Additionally, one can also consider another alternative description based on an antisymmetric field, the so-called massless Kalb-Ramond (KR) model \cite{duaw,ka1w,kaw}. Interestingly, the massive version of the latter describes spin-$1$ particles meaning that a degree of freedom discontinuity occurs at the zero mass limit. Regarding spin-$1$ particles, there is also an alternative dual description in terms of a symmetric tensor field \cite{d1w}. The latter also presents the previously mentioned degree of freedom jump in the massless limit. This is a topic of current research, see the following papers addressing this discontinuity for the (KR) model \cite{ana,meu} and also for the case of non-Abelian theories \cite{non}. \\
\indent Regarding the higher derivative spin-$0$ model studied in this paper, it can be related to the previously mentioned vector second-order derivative one through the master action technique \cite{d1w}. Interestingly, this higher derivative (UV) completion also contributes to overcoming the massless discontinuity presented by the parent model. This target model also has a reducible gauge symmetry structure with direct implications in the definition of the path integral formulation. Here, we intend to be the first ones to investigate these issues using a complementary Heisenberg operator quantization in an indefinite metric Hilbert space.  \\
\indent We extend the Kugo-Ojima-Nakanishi (KON) covariant operator formalism \cite{Nak1w,Nakqw} to quantize two Ostrogadskian systems, taking into account their generalized phase space structures \cite{ostrow,ostro2w}. More specifically, the one describing spin-$0$ particles by a vector field \cite{d1w} and the Bopp-Podolsky generalized electrodynamics \cite{pod1w}. We discuss a set of features related to the reducible structure of the local freedom \cite{batw} and a well-defined massless limit \cite{ana} for the first, and topics related to the correct definition of the positive norm subspace for the latter model, which is known to present transverse modes with a negative norm. The interacting regime is more involved since ghost particles can be created from an initial physical state.\\
\indent The (KON)  formalism is suitable to be applied for Lagrangian systems with a singular Hessian matrix. A set of auxiliary fields is properly included to provide a covariant model whose primary constraints are of second-class in the Dirac classification \cite{dirw}. In this manner, the Dirac brackets can be introduced to correctly project the model into the physical surface in which the constraints are valid in their strong form, eliminating quantization ambiguities. Therefore, considering the correspondence principle, the whole set of commutator's initial conditions can be obtained. From the operator equations of motion, the full commutator structure at unequal times can be provided. The quantization formalism is based on the Heisenberg description of an indefinite metric Hilbert space. This is a useful setting since the auxiliary fields as well as the negative norm spurious gauge field modes can be excluded from the physical subspace by an appropriate definition of the latter. This is achieved by a condition imposed by the auxiliary fields or even the BRST charges acting on the states, depending on the specific model considered.\\ 
 \indent This formalism has a relevant variety of applications ranging from Abelian and non-Abelian gauge theories to quantum gravity \cite{Nak1w}. It is possible to cite a variety of research associated with this formalism, such as the investigation of quantum gravity in tetrad formalism\footnote{For which the standard approaches generally fail.} \cite{ref1}, the definition of a complementary tool for BRST symmetry extensions \cite{ref2},  the discussion of mass generation mechanisms \cite{ref3}, in QCD confinement research \cite{conf} and to unveil formal aspects of QED$_4$ in the so-called non-linear t'Hooft gauge \cite{BRST,thooft}. One can also mention the application in topological condensed matter describing subtle aspects of the statistical interaction \cite{ann}. Interestingly, this approach can directly provide the exact quantization of the two-dimensional non-Abelian BF model \cite{Nak2Dw}.\\
 \indent This formalism also has a perturbative counterpart. The general structure of the Wightman functions can be inferred by a method that consists of extracting the truncated $n$ point functions from the $n$ point commutators by requiring the positive energy condition. For a review including applications in the perturbative quantization of $QED_4$, a toy one-loop model, string theory, and the two-dimensional quantum gravity, see \cite{Nakpertw}. Here, we provide a set of extensions of such formalism. Since both models investigated in this paper are of a higher derivative nature, one must consider the extended phase space and new fundamental Poisson brackets that characterize the Ostrogadskian systems. It has a direct influence on the set of initial conditions obtained by the correspondence principle. Moreover, regarding the higher derivative description of the spinless vector model, issues related to the reducible spurious sector and the necessity of adding two auxiliary fields are novelties introduced here. Then, in connection with the previously mentioned recent correlated achievements, we show how the Hilbert space structure behaves in the massless limit \cite{ana}.  Regarding the generalized QED$_4$, we explicitly show how the successful extension of the formalism indeed demands a higher derivative gauge fixing Lagrangian, being a full quantum completion to the previous efforts \cite{pimcaw} implicated by semi-classical considerations. Moreover, we show how to consistently define the positive-definite Hilbert subspace for the intricate case of the Bopp-Podolsky model presenting transverse massive negative norm modes. The interacting regime is also investigated, providing a set of formal discussions pointing out that a time-invariant physical subspace free of ghosts cannot be reached for this specific interaction. However, $S$-matrix unitarity is kept and a set of renormalization improvements associated with the presence of ghost modes are discussed and reviewed using different complementary approaches. In connection with a clarifying recent discussion on higher derivative theories, we establish a range of energies in which the external ghosts are absent, leading to a consistent probabilistic interpretation. Considering the achievements from the previous discussions, we are the first to introduce a higher derivative model with a specific interaction for which a time-invariant positive definite Hilbert subspace can be consistently established due to an additional discrete symmetry. \\
\indent The work is organized as follows. In Sec. \ref{q}, the Lagrangian structure and the pertinent auxiliary fields for the higher derivative spin-$0$ model are provided. Later, at Sec. \ref{p}, the complete set of equations of motion for the system is derived. The Sec. \ref{4} is devoted to characterizing the phase space of the system as well as obtaining the primary constraint structure which is second-class due to the presence of the previously mentioned auxiliary fields. In Sec. \ref{5}, the correspondence principle and the complete set of commutators are derived. Moreover, the definition of the physical subspace and its massless limit are investigated. The Sec. \ref{6} introduces the Lagrangian of the higher derivative electrodynamics, its auxiliary fields, equations of motion, primary constraints, and the Dirac Brackets. Then, in Sec. \ref{7} the full commutator structure is derived, as well as the definition of a physical Hilbert space capable of avoiding spurious gauge projections and also the massive transverse modes with negative norms. It is demonstrated that the Hamiltonian becomes positive-definite in this subspace. The Sec. \ref{8} discusses the theory in the presence of a fermionic interaction. A set of formal discussions are presented regarding unitarity and probabilistic interpretation. Although ghosts generally cannot be decoupled in this regime, we present an alternative complementary analysis, based on the optical theorem, of their role in decreasing the divergence degree for the $n$-point functions. After these efforts, we present a specific interacting higher derivative model compatible with establishing a time-invariant positive subspace. Finally, we conclude in Sec. \ref{9}. The metric signature $(+,-,-,-)$ is used throughout this work.

\section{On the Lagrangian structure and auxiliary fields for a dual model describing spin-$0$ particles }\label{q}

\indent The Lagrangian describing a spin-$0$ particle using a vector field embedded in a higher derivative structure  reads \cite{d1w}

\begin{align} {\mathcal{L}} =\frac{1}{2}\Big[\partial_\beta\big(\partial_\mu B^\mu\big)\partial^\beta\big(\p_\mu B^\mu\big)-m^2\big(\p_\mu B^\mu\big)^2     \Big]+\epsilon^{\mu \nu \rho}\p_\nu \phi_\rho\big(\Box+m^2\big)B_\mu+\partial^\mu \Omega\big(\Box+m^2\big)\phi_\mu  \nonumber \\              \end{align}
in which the auxiliary fields are explicitly highlighted. To establish a scalar Lagrangian, both auxiliary fields $\phi_\mu(x)$ and $\Omega(x)$ must be pseudo vector and pseudo scalar fields, respectively. We choose to work in $D=2+1$ dimensions since, in this case, just two auxiliary fields are necessary to provide a second-class system in compliance with the correspondence principle.
Therefore, the model has well-defined Dirac brackets suitable for developing the quantization procedure. We will explicitly prove it in the Sec. \ref{4}.\\
\indent Up to the application of the Klein-Gordon operator, the theory would be related to the massless Kalb-Ramond model, leading to no degrees of freedom for a planar quantum field theory. Since the presence of this operator, implying an Ostrogadskian structure, leads to a theory with a propagating degree of freedom and also complies with a well-defined massless limit, we find it relevant to study this toy model as a prelude to the generalized QED$_4$. \\
\indent The auxiliary fields are responsible for breaking the local symmetry and providing a gauge condition valid in all Hilbert space. This is a demand due to the singular nature of the theory composed just by the gauge vector field $B_\mu(x)$. In the absence of the gauge fixing Lagrange multipliers, the theory presents the following local freedom
\bea  B_\mu(x)\to  B_\mu(x)+ \epsilon_{\mu \nu \sigma}\p^\nu \Lambda^\sigma(x)\eea 
with $\Lambda_\sigma(x)$ being a pseudo vector field representing the symmetry parameter. This transformation has a reducible structure presenting the zero modes
\bea \Lambda_\alpha(x)=\p_\alpha \rho(x)  \eea
in which $\rho(x)$ is an arbitrary pseudo scalar field.\\
\indent As mentioned, the presence of the whole set of auxiliary fields breaks the symmetry through the condition  \footnote{with the definition $F_{\mu \nu}^{B}(x)\equiv \p_\mu B_\nu(x)-\p_\nu B_\mu(x)$} \cite{ostrow}   \bea \big(\Box+m^2\big)\big(\epsilon^{\mu \nu \beta}F_{\mu \nu}^{B}(x)-2\p^\beta \Omega \big)=0\eea
Due to the existence of the zero modes, the extra pseudo scalar auxiliary field is demanded to eliminate the freedom associated with any first-class quantities \cite{Nak1w}. Although performed in $D=2+1$, this process can be generalized for any dimension. We focus on this particular one since, in this case, just one extra auxiliary field is necessary, defining a straightforward generalization of the well-known (KON) quantization. In the reference \cite{2024}, an analogous quantization process is generalized for $D$-dimensions for the case of a dual description of a spin $1$ model. Finally, beyond this point associated with the reducible nature of the symmetry, the system also present a higher derivative structure, implying the necessity of introducing an extended phase space.   \\ 

\section{On the equations of motion}\label{p}
\indent The model's equations of motion are given below

\begin{align} -\p^\mu\big(\Box+m^2\big)\big(\p_\nu B^\nu(x)\big)+\big(\Box+m^2\big)\epsilon^{\mu \nu \rho}\p_\nu \phi_\rho(x)=&0\ ,\nonumber \\
\big(\Box+m^2\big)\p_\mu \phi^\mu=&0\ ,\nonumber \\
-\epsilon^{\mu \nu \rho}\p_\nu \big(\Box+m^2\big)B_\mu(x)+\p^\rho \big(\Box+m^2\big)\Omega(x)=&0 \label{eqq}\end{align}
\indent Taking the divergence of the first equation yields  
 \bea \Box\big(\Box+m^2\big)\p_\mu B^\mu(x)=0\eea
 \indent This equation has a deep analogy with the two pole structure of the Bopp-Podolsky generalized electrodynamics, despite the facts that there is no negative residue and the pole field is a scalar combination $\p_\mu B^\mu(x)$. \\
\indent From the divergence of the last equation of \eqref{eqq}, it is possible to derive the pole structure for the auxiliary field $\Omega(x)$ 
\bea \Box\big(\Box+m^2\big)\Omega(x)=0\eea
\indent Applying the differential operator $\epsilon_{\mu \nu \alpha}\p^\nu$ on the first equation and using the auxiliary vector field transverse nature, implies  the equation
\bea \Box\big(\Box+m^2\big) \phi^\mu(x)=0\eea
defining the pole structure of the vector auxiliary field.

\section{On the extended phase space structure}\label{4}

\indent After this digression, we introduce the generalized Ostrogadskian canonical momenta structure which, for a fourth derivative model, is defined below
\begin{align} p_\Phi(x) \equiv& \Big[\frac{\partial{\mathcal{L}}}{\p(\p_0 \Phi(x))}-2\p_i\Big(\frac{\partial{\mathcal{L}}}{\p(\p_i\p_0 \Phi(x))}\Big)-\p_0\frac{\partial{\mathcal{L}}}{\p(\p_0\p_0 \Phi(x))} \Big] \nonumber \\
\pi_\Phi(x)\equiv &\frac{\partial{\mathcal{L}} }{\p(\p_0\p_0 \Phi(x))}\nonumber \\
\end{align}
with $\Phi(x)$ being a general notation for all fields appearing in the Lagrangian.\\
\indent Then, for the specific system studied by us, the generalized momenta definition is expressed below

\begin{align} &\pi_\mu^B(x)=\big(\ddot B^0(x)+\p_i\dot B^i(x)\big)\delta^0_\mu+\epsilon_{\mu \nu \rho}\p_\nu \phi_\rho(x),\nonumber \\ &p^B_\mu(x)=-m^2\big(\dot B^0(x)+\p_iB^i(x)\big)\delta^0_\mu+2\p^k\p_k\big(\dot B^0(x)+\p_iB^i(x)\big)\delta^0_\mu\nonumber \\ &-2\p^k\big(\ddot B^0(x)+\p_i\dot B^i(x)\big)\delta_k^\mu,\nonumber -\big(\dddot B^0(x)+\p_i \ddot B^i(x)\big)\delta^0_\mu-\epsilon_{\mu \nu \rho}\p^\nu \ddot \phi_\rho(x), \nonumber \\
 &p^{\phi}_\mu(x)=\epsilon^{k0\mu}\big(\Box+m^2\big)B_k(x)-\p^\mu \ddot \Omega(x) \ ,\
 \pi^{\phi}_\mu(x)=\p_\mu \Omega(x)\ , \nonumber \\
&p^{\Omega}(x)=\big(\Box+m^2\big)\phi_0(x)\ ,\
 \pi^\Omega(x)=0\label{mom}\end{align}
\indent In order to define the nature of the primary constraints, we first introduce the fundamental Poisson brackets in this extended phase space $\Big(\Phi(x),\dot \Phi(x),p^{\Phi}(x),\pi^{\Phi}(x)\Big)$

\begin{align}&\Big\{ \Phi(x),p^{\Phi}(y)\Big\}={\cal{I}}\delta^2(\vec x-\vec y)\ , \ \Big\{ \dot \Phi(x),\pi^{\Phi}(y)\Big\}={\cal{I}}\delta^2(\vec x-\vec y)                \label{phase}         \end{align}
with ${\cal{I}}$ denoting the identity written in terms of the specific field tensor structure of a given field theory.\\
\indent The primary constraints are the momentum definitions generating non-dynamical relations between the phase space degrees of freedom. They are the following $\pi_j^B(x)=\epsilon^{j0k}\dot \phi_k(x)+\epsilon^{jl0}\p_l\phi_0(x)$, $\pi^{\phi}_\mu(x)=\p^\mu \Omega(x) $ and $\pi^\Omega(x)=0$. Then, it proves that the correct introduction of the complete set of auxiliary fields indeed leads to a full second-class system from the beginning, even in this higher derivative scenario. The constraints can be suitably grouped as
 \begin{align}
      \mathcal{C}^{(1)}_\mu(x) &\equiv  \big(-\pi^j_B(x)-\epsilon^{ij}\dot \phi_j(x)+\epsilon^{ij}\p_j\phi_0(x)\big)\delta_i^\mu+\pi^\Omega(x) \delta_0^\mu \approx 0 \nonumber \\  \mathcal{C}^{(2)}_\mu(x)& \equiv \pi^{\phi}_\mu(x)-\p^\mu \Omega(x)\approx 0 \end{align}
being associated with the following inverse matrix of constraints
 \bea \mathcal{G}_{IJ}^{\mu \nu}(x,y)=\{\mathcal{C}^I_\mu(x),\mathcal{C}^J_\nu(y)\}^{(-1)}=\left(\begin{array}{ccc}
	0& -\epsilon_{ij}\delta^i_\mu \delta^j_\nu+  \delta_\mu^0 \delta_\nu^0     \\
	\epsilon_{ij}\delta^i_\mu \delta^j_\nu-\delta_\mu^0 \delta_\nu^0 & 0  \\
\end{array}\right)\delta^2(\vec x-\vec y)  \eea
enabling one to establish the Dirac brackets, the key object to provide the system's quantization
\bea \Big\{F(x),G(y) \Big\}_D=\Big \{F(x),G(y) \Big \}-\int d^3z\ d^3w \Big \{F(x),
 \mathcal{C}^I_\mu(z)\Big \}\mathcal{G}_{IJ}^{\mu \nu}(z,w)\Big \{ \mathcal{C}^J_\nu(w),G(y)\Big \}\eea
 for which the constraints are valid in their strong form.

\section{From the commutator initial conditions to its final form at unequal times }\label{5}

\indent From the Dirac brackets, the correspondence principle can be established yielding the following set of non-vanishing commutators
\begin{align} \Big[B_\mu(x),p_B^\nu(y)\Big]_0&=i\delta^\nu_\mu \delta^2(\vec x-\vec y) \ , \  \Big[\dot B_\mu(x),\pi_B^\nu(y)\Big]_0=i\delta^\nu_\mu \delta^2(\vec x-\vec y) \nonumber \\   \Big[\Omega(x),p^\Omega(y)\Big]_0&=i\delta^2(\vec x-\vec y)\ , \ \quad  \ \Big[\phi_\mu(x),p_\phi^\nu(y)\Big]_0=i\delta^\nu_\mu \delta^2(\vec x-\vec y)\nonumber \\ \Big[\p_0\phi_\mu(x),\pi_\phi^\nu(y)\Big]_0&=i\delta^\nu_\mu \delta^2(\vec x-\vec y)\nonumber         \end{align}
leading to a set of commutator's initial conditions due to the momentum definition and the automatic imposition of the primary constraints

\begin{align} \Big[\dot B_0(x),\ddot B_0(y)\Big]_0&=i\delta^2(\vec x-\vec y)\ , \
\Big[\dot B_j(x),\epsilon^{ik}\dot \phi_k(y)\Big]_0=-i\delta_j^i\delta^2(\vec x-\vec y),\nonumber \\
\Big[B_0(x),\dddot B_0(y)\Big]_0&=-i\delta^2(\vec x-\vec y)\ , \
\quad \Big[\phi_0(x),\ddot \Omega(y)\Big]_0=-i\delta^2(\vec x-\vec y),\nonumber \\
 \Big[\dot \phi_0(x),\dot \Omega(y)\Big]_0&=i\delta^2(\vec x-\vec y)\ , \
\ \Big[\phi_i(x),\epsilon^{jk}\ddot B_k(x)\Big]_0=i\delta_i^j\delta^2(\vec x-\vec y)\end{align}

\indent The first commutator to be addressed here is associated with the auxiliary scalar field. Since it obeys  \bea \Box \big(\Box+m^2\big)^x\Big[\Omega(x),\Omega(y)\Big]=0\eea
one can use the properties of the distributions below \cite{Nak1w}
\begin{multline}
\Box \Delta(x-y; s) = -s\Delta(x-y; s), \quad \Delta(x-y; s)\Bigr|_{\substack{0}} = 0, \quad \dot \Delta(x-y; s)\Bigr|_{\substack{0}} = \delta^2(\vec x-\vec y),  \\ 
\big(\Box+s\big) E(x-y; s)=\Delta(x-y; s), \ \ E(x-y; s)\Bigr|_{\substack{0}}=0, \ \ \dddot E(x-y; s)\Bigr|_{\substack{0}}=\delta^2(\vec x-\vec y). 
\label{CauchyGreenw}
\end{multline}
representing their initial conditions, and showing that the general solution must be of the form
\bea   \Big[\Omega(x),\Omega(y)          \Big]=a \Delta(x-y,m)+b\Delta(x-y,0)       \eea

Owing to the canonical momenta definition \eqref{mom}, one can obtain $\dot \Omega(x) $ in terms of $\pi_0^\phi(x)$. Then, considering the general Ostrogadskian phase space structure \eqref{phase}, it is possible to derive the initial condition $\Big[\dot \Omega(x),\Omega(y)          \Big]_0=0$
implying $a+b=0$. Since $p^\phi_0(x)$ and $\pi^\phi_0(x)$ are not phase space conjugate variables, we also have $\Big[\dot \Omega(x),\ddot \Omega(y)          \Big]_0=0$, 
leading to $a=0$ and, consequently, a vanishing commutator at unequal times
\bea \Big[\Omega(x),\Omega(y)          \Big]=0\eea 
\indent Following similar reasoning, one obtains  $\Big[\Omega(x),B_\nu(y)\Big]=0$.

Now, we derive the commutator between the gauge and vector auxiliary field. As already mentioned, this and the previous commutator are the key ones to understand how a consistent definition of the positive Hilbert space can be established. The general form is given below,
\begin{align} \Big[B_\mu(x), \phi_\rho(y)\Big]&=i\epsilon_{\mu \rho \nu}\p^\nu\big(a \Delta(x-y,m^2)+b\Delta(x-y,0)\big) \end{align}
considering the pseudo-vector nature of $\phi_\mu(x)$ and its transverse condition implicated by the operator equations of motion. This commutator complies with the fact that $\phi_\rho(x)$ lies in the kernel of the $\Box\big(\Box+m^2\big)$ operator and obeys the gauge condition $(\Box+m^2)\p_\mu \phi^\mu(x)=0$. From the initial conditions
$\Big[B_j(x),\ddot \phi_m(y)\Big]_0=i\epsilon_{mj}\delta^2(\vec x-\vec y)$
the relation $a=-b$ with  $a=\frac{i}{m^2}$ is achieved, specifying the full structure at unequal times
\
\bea \Big[B_\mu(x), \phi_\rho(y)\Big]=\frac{i}{m^2}\epsilon_{\mu \rho \nu}\p^\nu\big( \Delta(x-y,m^2)-\Delta(x-y,0)\big)\eea
\indent It means that just the non-physical transverse part of $B_\mu(x)$ has a non-zero commutator with $\phi_\mu(x)$. It furnishes a background to further deriving the subsidiary condition defining the positive-definite Hilbert subspace. \\
\indent In order to achieve this objective, the commutator between the different auxiliary fields must be obtained.  Owing to Lorentz covariance, the general form for the mentioned commutator is given below
\bea \Big[\phi_\rho(x),\Omega(y)\Big]=\p_\rho\big(a\Delta(x-y,m^2)+b\Delta(x-y,0)\big)\eea
since both field operators are in the kernel of $\Box\big(\Box+m^2\big)$ and the scalar one is coupled to the longitudinal part of $\phi_\mu(x)$. Then, the initial conditions $ \Big[\phi_0(x),\ddot \Omega(y)\Big]_0=-i\delta^2(\vec x-\vec y)$
as well as the distribution properties imply $a=-b$ and $a=-\frac{i}{m^2}$, leading to

\bea \Big[\phi_\rho(x),\Omega(y)\Big]=-\frac{i}{m^2}\p_\rho\big(\Delta(x-y,m^2)-\Delta(x-y,0)\big)\eea
\indent It is worth mentioning that the last two commutators present well-defined massless limits since $\Delta(x-y,m^2) =\big(\Delta(x-y,0)-m^2E(x-y,0)+...\big)$ for $m\to 0$. This is an important distinguishing feature since there are other dual models with reducible local symmetries presenting a kind of DVZ-like discontinuity in their massless limits. Concretely, one can cite the Kalb-Ramond model describing scalar particles or spin-$1$ ones in terms of an anti-symmetric tensor field in the massless/massive phases, respectively.\\
\indent The derivation of the commutator between the auxiliary vector fields is the final goal to achieve a complete description of the auxiliary sector.  Considering the gauge fixing equation and the initial conditions, one obtains
\begin{align} \Big[\phi_\mu(x),\phi_\nu(y)\Big]=0\end{align}
defining the zero norm nature of such field operators.\\
\indent The last step before defining the physical Hilbert space is the derivation of the vector gauge field commutator. Considering its equations of motion and the set of initial conditions, an analogous procedure yields
\bea \Big[B_\mu(x),B_\nu(y)\Big]=\frac{i}{m^4}\p_\mu \p_\nu\big(\Delta(x-y,m^2)-\Delta(x-y,0)\big)+\frac{i}{m^2}\p_\mu \p_\nu E(x-y,0)\eea
\indent It furnishes the exact scalar commutator for the pole operators
\bea \Big[\p^\mu B_\mu(x),\p^\nu B_\nu(y)\Big]=i\Delta(x-y,m^2)\eea
\indent It is worth mentioning that this structure defines a spin-$0$ particle even at the massless limit. An alternative check to this conclusion is the fact that the inter-particle potential derived for this model in \cite{gabw} is well-defined at this limit. This theory is a higher derivative version of the second-order vector spin-$0$ one \cite{d1w}, obtained through the master action technique. Since the latter loses its particle content at $m\to 0$, it indicates that using a higher derivative structure may avoid these discontinuities. A full discussion on this theme is provided in the appendix, displaying the complete Hamiltonian analysis of the model's Ostrogadskian phase space.\\
\indent Having established the commutator structure for all the fields, one can
derive a suitable subsidiary condition to avoid the emergence of the auxiliary fields in the positive Hilbert subspace.\\
 Therefore, a good subsidiary condition to define the positive semi-definite Hilbert subspace is
\begin{equation}
\phi_\mu^+(x) | \text{phys} \rangle = 0, \quad \forall | \text{phys} \rangle \in \mathcal{V}_{\text{phys}}.
\label{conditionBfreefieldw}
\end{equation}
in which $\phi_\mu^+(x)=\phi_{\mu}^{+(m)}(x)+\phi_{\mu}^{+(0)}(x)$ denotes the sum of the positive frequency parts of the massive and massless solutions of the $\phi(x)$ field equations of motion.  According to the whole set of the previously derived commutators, this 
definition eliminates the spurious non-positive norm gauge field projections from the positive-definite metric Hilbert subspace  
\bea    \mathcal{H}_{phys}=\frac{\mathcal{V}_{phys}}{\mathcal{V}_{0}}          \eea
defined as the completion of the quotient space above. The zero norm states associated with the action of the auxiliary fields on the vacuum, spanning the space $\mathcal{V}_{0}$, are also suitably avoided by this definition.\\
\indent It is straightforward to show that the scalar operator $\p_\mu B^\mu(x)$ commutes with the vector auxiliary field, fulfilling the subsidiary condition. It means that the negative frequency part of this field creates a scalar particle with a positive norm when acting on the vacuum state, defining a physical state.  Therefore, the present developments provide an extension of the (KON) formalism for systems with reducible gauge symmetry structure and of fourth order in derivatives.\\

\section{The Bopp-Podolsky Higher Derivative Electrodynamics}\label{6}

\indent The Lagrangian for the higher order electrodynamics reads \cite{pod1w,pod2w,pod3w}
\bea {\mathcal{L}}=-\frac{1}{4}F_{\mu \nu}F^{\mu \nu}+\frac{1}{2m^2}\p_\lambda F^{\alpha \lambda}\p^{\rho}F_{\alpha \rho}+\p^\mu B\Big(\frac{\Box}{m^2}+1\Big)A_\mu\label{eom0} \eea
\indent We consider a higher derivative structure for the gauge fixing sector\footnote{Breaking the local freedom $A_\mu(x) \to A_\mu(x)+\p_\mu \beta(x)$, with $\beta(x)$ being an arbitrary scalar.} due to two reasons. First, as we are going to see, it is the most general condition compatible with a pole equation for the vector field in the physical subspace \cite{pimcaw}. The other is; considering the previously introduced Ostrogadskian phase space structure, this higher-order term contributes to generating a set of non-vanishing generalized momenta responsible for turning the system into a second-class one, enabling the establishment of the correspondence principle through the Dirac Brackets. This is one of the fundamental underlying principles of the covariant operator formalism \cite{Nak1w,Nakpertw}.\\
\indent The equations of motion are the following 
\begin{align} \big(\Box+m^2\big)\p^\mu A_\mu&=0\ , \quad \big(\Box+m^2\big)\big(\p^\nu F_{\nu \mu}+\p_\mu B\big)=0\ , \quad \Box\big(\Box+m^2\big)B=0\label{eom} \end{align}
\indent From the previous relations, the vector field equation can be properly decoupled
\bea \Box^2 \big(\Box+m^2\big)A_\mu(x)=0 \eea
\indent According to \eqref{mom}, the phase space variables in this Ostrogadskian \cite{ostrow,ostro2w} system reads
\begin{align} p_\alpha(x)&=-F_{0\alpha}(x)-\frac{1}{m^2}\big(\p_k\p_\lambda F^{0\lambda}(x)\delta^k_\alpha-\p_0\p_\lambda F_\alpha^{\ \lambda}-\p_\alpha \p_0 B\big),\nonumber \\
\pi_\alpha(x)&=\frac{1}{m^2}\big(\p_\lambda F^{0\lambda}(x)\delta^0_\alpha-\p_\lambda F_{\alpha}^{\ \lambda}(x)\big)+\frac{1}{m^2}\p_\alpha B(x),\nonumber \\
p_B(x)&=(\frac{\Box}{m^2}+1)A_0\ , \quad \pi_B(x)=0
\end{align}

\indent The momenta definition furnishes two constraints which are of second-class \cite{dirw}, see \cite{pimcaw}
\bea \mathcal{C}^{(1)}\equiv \pi_0(x)-\frac{1}{m^2}\dot B(x)\approx 0 \quad , \quad \mathcal{C}^{(2)}\equiv \pi_B(x)\approx 0\eea
implying that the gauge fixing sector has the required structure to develop the quantization process. The inverse of the constraint matrix reads
   \bea \mathcal{G}^{IJ}(x,y)=\{\mathcal{C}^I(x),\mathcal{C}^J(y)\}^{(-1)}=\left(\begin{array}{ccccc}
	0& +m^2    \\
	-{m^2} & 0       \\
	
\end{array}\right)\delta^3(\vec x-\vec y)  \eea
which can be used to build the Dirac brackets, the fundamental object to define the quantization process by the correspondence principle
\bea  \{F(x),G(y)\}_D=\{F(x),G(y)\}-\int d^3z d^3w \{F(x),\mathcal{C}_I(z)\} \mathcal{G}^{IJ}(z,w)\{\mathcal{C}^J(w),G(y)\}\label{diracb}\eea
Then, a well-defined bracket for which the constraints are valid in the strong form is achieved.

\section{On the commutator structure}\label{7}

\indent In order to define all the field commutators at unequal times, one must consider the equations of motion and the initial conditions provided by the correspondence principle. The extended phase space for a system of fourth order in derivatives is defined in the following manner $\epsilon_\mu\equiv \big(A_\mu(x), \dot A_\mu(x), B(x), \dot B(x), p_\alpha(x),\pi_\alpha(x), p_B(x), \pi_B(x)  \big)$ and the fundamental Poisson brackets structure is the one defined in \eqref{phase}. With this knowledge, the Dirac brackets \eqref{diracb} can be established. Then, considering the correspondence principle, the following set of initial conditions can be derived
\begin{align}  \Big[\dot A_i(x),\ddot A^j(y)\Big]_0&=im^2\delta_i^j\delta^3(\vec x-\vec y)\ ,\quad \Big[A_0(x),\ddot B(y)\Big]_0= im^2\delta^3(\vec x-\vec y),\nonumber \\
  \Big[A_\mu(x),\dot A^\nu(y)\Big]_0&=0\ , \quad \quad \quad \quad \quad \quad \quad  
  \Big[A_i(x),\dddot A^j(y)\Big]_0=-im^2\delta_i^j\delta^3(\vec x-\vec y)\label{aw}              \end{align}

\indent Now, inferring the general structure of the commutator for the gauge and the auxiliary field, demands the use of the operator equations of motion to obtain a condition valid at unequal times
\bea \Box^y\big(\Box+m^2\big)^y\Big[A_\mu(x),B(y)\Big]=0\eea

\indent Then, owing to Lorentz covariance, we establish the most general commutator in the kernel of the above equation
\bea \Big[A_\mu(x),B(y)\Big]=a\p_\mu \Delta(x-y;0)+b\p_\mu \Delta(x-y;m^2)\eea
in which $a$ and $b$ are undetermined constants. In this manner, considering the initial condition 
$\Big[A_0(x),\ddot B(y)\Big]_0=im^2\delta^3(\vec x-\vec y)$
 one concludes that $a=-b$ and $a=i$, leading to
 \bea \Big[A_\mu(x),B(y)\Big]=i\big(\p_\mu \Delta(x-y;0)-\p_\mu \Delta(x-y;m^2)\big) \label{comut1}\eea

The commutator structure above allows us to conclude that the auxiliary field does not commute with the spurious longitudinal projections of the gauge field. \\
\indent Considering a similar approach, one can also derive
\bea \Big[B(x),B(y)\Big]=0\eea 
expressing the zero-norm character of the auxiliary field.\\
\indent Taking into account the equations of motion for the vector field, and the initial conditions, an analogous procedure lead to the full commutator for the vector field
\begin{align} \Big[A_\mu(x),A_\nu(y)\Big]&=-i\big(\eta_{\mu \nu}\Delta(x-y,0)-\p_\mu\p_\nu E(x-y,0)\big)\nonumber \\&+i\big(\eta_{\mu \nu}\Delta(x-y,m^2)+\frac{1}{m^2}\p_\mu\p_\nu \Delta(x-y,m^2)\big)\nonumber \\&-i\frac{1}{m^2}\p_\mu \p_\nu \Delta(x-y,0)\label{comut2}\end{align}
\indent Considering the fact that the simple and double pole massive Pauli-Jordan distributions  tend to zero at  $m \to \infty$, the Maxwellian form for the commutator is recovered at this limit \cite{Nak1w} 
\bea \Big[A_\mu(x),A_\nu(y)\Big]=-i\big(\eta_{\mu \nu}\Delta(x-y;0)-\p_\mu\p_\nu E(x-y;0)\big)  \eea
\indent Now, considering the commutator structure of \eqref{comut2}, the subsidiary condition can be defined. Let's try to apply the same one used to characterize the positive semi-definite Hilbert subspace of QED$_4$
\begin{equation}
B^+(x) | \text{phys} \rangle = 0, \quad \forall | \text{phys} \rangle \in \mathcal{V}_{\text{phys}}.
\end{equation}
with $B^+(x)=B^{+(0)}(x)+B^{+(m)}(x) $ being the sum of massive and massless positive frequency parts of the auxiliary field operator. \\
\indent Although this condition prevents spurious gauge field projections from appearing in the physical subspace, it does not eliminate transverse massive negative norm states associated with the algebra \eqref{comut2}.\\
\indent This feature is not a failure of the $B$ field approach, this model indeed presents this kind of ghost particle as well as, for example, some higher derivative gravity models \cite{shapw}. However, we can generalize this approach to separate the positive semi-definite Hilbert subspace. Then, we consider two conditions \footnote{Here, $B^{+(0)}(x)\equiv (\Box+m^2)B^+(x)$, with the $+$ label denoting the positive frequency part of the field.}
 \begin{equation}
B^{+(0)}(x) | \text{phys} \rangle = 0,\quad \Box^2 A_\mu^+(x) | \text{phys} \rangle = 0 \quad \forall | \text{phys} \rangle \in \mathcal{V}_{\text{phys}}.
\label{2}
\end{equation}

\indent First, we notice that the two conditions are compatible. Moreover, the auxiliary field $B^{+(m)}(x)$ and the whole massive pole contribution to $A_\mu(x)$ are necessarily outside the physical subspace due to commutations relations \eqref{comut1} and \eqref{comut2}. The longitudinal spurious contribution from the massless sector is also excluded due to the condition generated by the B-field. Therefore, we can establish a well-defined positive-definite Hilbert subspace even for the Bopp-Podolsky model as the Cauchy completion of 
\bea \mathcal{H}_{\text{phys}}=\frac{\mathcal{V}_{\text{phys}}}{\mathcal{V}_0}\eea
 with ${\mathcal{V}_0}$ representing the zero norm subspace. This definition is time-invariant for this free theory. However, when including an interaction, one cannot exclude the massive ghosts, since they can be produced by time evolution.  \\

 \subsection{On the positive-definite Hilbert subspace} 

 \indent The quantum vector field can be expressed in terms of creation and annihilation operators as \footnote{In natural units.} 
\bea  A_\mu(x)=\int \frac{d^3p}{(2\pi)^{3/2}}\Big[  \sum_{\lambda}\frac{1}{\sqrt{2|\vec p|}}\big(e^{-ipx}a_\lambda(\vec p)\epsilon_\mu^\lambda(\vec p)+e^{ipx}a_\lambda^\dagger(\vec p) \epsilon^{*\lambda }_\mu(\vec p)\big)+ \\ \nonumber
\sum_{\sigma}\frac{1}{\sqrt{2\sqrt{|\vec p|^2+m^2}   }}\big(e^{-i\bar px}\bar a_\sigma(\vec {\bar p}) \chi_\mu^\sigma(\vec {\bar p})+e^{i\bar px}\bar a_\sigma^\dagger(\vec {\bar p}) \chi^{*\sigma }_\mu(\vec{\bar p})\big)\Big]  \eea
with $p_\mu=(\sqrt{\vec p^2},\vec p)$, $\bar p_\mu=(\sqrt{\vec p^2+m^2},\vec p)$ and $[a^\lambda(\vec p),(a^{\lambda'}(\vec q))^\dagger]=-[\bar{a}^\lambda(\vec p),(\bar{a}^{\lambda'}(\vec q))^\dagger]=\eta^{\lambda \lambda'}\delta^3(\vec p-\vec q)$.\\
\indent This field operator structure is in compliance with equation \eqref{comut2} if the following polarization sums,
\bea   \eta_{\lambda \lambda'} \epsilon^\lambda_\mu(\vec p)\epsilon^{*\lambda'}_\nu(\vec p)=- \theta_{\mu \nu}( p)-\frac{p_\mu p_\nu}{m^2} \label{pol1} \eea
\bea   \eta_{\lambda \lambda'}\chi^\lambda_\mu(\vec{\bar p})\chi^{*\lambda'}_\nu(\vec{\bar p})=-g_{\mu \nu}+\frac{\bar p_\mu \bar p_\nu}{m^2} \label{pol2} \eea

with $\theta_{\mu \nu}=\eta_{\mu \nu}-\frac{p_\mu p_\nu}{p^2}$, are defined. Due to the presence of negative norm states, the completeness relation should be written as $\operatorname{I}=\sum_X \int d\Pi_X |X\rangle\langle X|-\int d\Pi_{X'}\sum_{X'} |X'\rangle\langle X'|$ representing a sum over the discrete labels and an integral over the continuous ones $d\Pi_x=\Pi_{i \in X}\frac{d^3p_j}{(2\pi)^32E_j}$ present in the eigenstates of the asymptotic Hamiltonian. Here, $X$ denotes all asymptotic particle labels contained in a given Fock state of positive norm and $X'$ has the same interpretation regarding the negative norm states. This structure indeed ensures that $\operatorname{I}|\psi\rangle=|\psi\rangle$, with $|\psi\rangle$ being an arbitrary state. \\
\indent Owing to the equation \eqref{comut2}, for a given specific frame $p_\mu=(p,0,0,p)$, the physical modes with positive norm have the following commutator
\bea \Big[\tilde{a}_i(p), \tilde{a}_j^\dagger(q)\Big]=\delta_{ij}(2\pi)^3\theta(p_0)\delta(p^2) \delta^4(p-q) \eea  with $i=1,2$ denoting spatial coordinates. They are associated with the alternative setting of the positive and negative frequency parts of the vector field \cite{Nak1w}
\bea    A_\mu^{(+)}(x)=\int d^4p\big( \tilde{a}_\mu(p)e^{-ip.x} +a'_\mu(p)e^{-ip.x}       \big) \ ; \ A_\mu(x)=A_\mu^{(+)}(x)+A_\mu^{(-)}(x)    \eea
The physical modes previously highlighted are associated with the annihilation operators as $\tilde{a}_\mu(p)\equiv \sum_\lambda a^\lambda(\vec p)\epsilon^\lambda_\mu(\vec p)\theta(p_0)\delta(p^2)\sqrt{2p_0}$. The operator $a_\mu'(p)$ is related to the massive sector.\\
\indent It is worth mentioning that the generalized QED$_4$ has a Hermitian Hamiltonian operator $H=H^\dagger$, ensuring  generalized unitarity \cite{Nak1w}. However, it is not positive-definite as a consequence of its Ostrogadskian structure.  Despite this fact, considering the definition of the physical Hilbert subspace \eqref{2}, one can show that the matrix elements 
\bea \langle \text{phys} A|:H:|B\text{phys}\rangle=\langle \text{phys}A|\sum_{i=1}^2\int \frac{d^3k}{(2\pi)^3}E(k) a_ia^\dagger_i|B\text{phys}\rangle \label{hamiltonian} \eea
are positive-definite with $E(k)=|\vec k|$. The labels $A$ and $B$ are employed to identify an arbitrary pair of different physical states. Therefore, Ostrogadskian instabilities are absent in the physical subspace of the free theory. We have defined $a_i(\vec p)\equiv \sum_\lambda a^\lambda(\vec p)\epsilon^\lambda_i(\vec p)$, with $i=1,2$ denoting the transverse spatial coordinates.

\section{Massive ghosts in an interacting theory}\label{8}

\indent A natural question that arises regards the possibility of consistently separating the physical Hilbert space in an interacting theory. In order to address this point, let's add to the free Bopp-Podolsky Lagrangian the following coupling term 
$A_\mu(x) J^\mu(x)$, written in terms of a fermionic current \footnote{$\psi(x)$ is the spinor field representing the electrons and positrons.} $J_\mu(x)=\bar \psi(x)\gamma_\mu \psi(x)$, and also the kinetic term for the matter field. Then, the commutation relations between the gauge field and the B-field are kept since $\p_\mu J^\mu(x)=0$. From the operator equations of motion \eqref{eom0} and \eqref{eom} in the presence of this interaction, one concludes that the $B(x)$ field remains free. Its commutator with the current is obtained through the operator equations of motion as 
\bea \Big[B(x),J_\mu(y)  \Big]=\Big[B(x), \big(\Box+m^2\big)\Box A_\mu(y) \Big]=0 \eea
using the information from \eqref{comut1}. Therefore, the current can be regarded as a gauge invariant operator, as it should be.\\
\indent It is possible to show that the asymptotic fields are still a combination of massive and massless excitations since the vacuum-corrected Feynman propagator has the form  \footnote{We are highlighting just its transverse physical part.}  \cite{pod7w}
\bea  P_{\mu \nu}(p^2)=\frac{im^2\ \theta_{\mu \nu}}{p^2(p^2-m^2-m^2\pi^R(p^2))}\label{prop} \eea
with  $\pi^R(p^2)$ denoting the scalar part of the complete polarization tensor structure. It can be renormalized to fulfill the condition $\pi^R(0)=0$ ensuring a positive unitary norm for the massless pole. On the other hand, the massive pole is renormalized by the radiative correction. Then, the asymptotic algebra is again of the form \eqref{comut2} with renormalized parameters. If the gauge field mass lies above the particle production threshold $m>2m_e$, with $m_e$ denoting the electron mass, the massive pole presents an imaginary part \footnote{$\alpha$ is the fine structure constant.} $ \gamma=m^2\operatorname{Im}\Big( \Pi^R(m^2)\Big)\approx \frac{ \alpha\ m^2}{3}\sqrt{1-\frac{4m^2_e}{m^2}}\big( 1+\frac{2m^2_e}{m^2} \big)>0$ with the right sign to define the so-called merlin mode structure for the massive pole \cite{grav}. However, we do not consider this last setting for the mass parameters and will follow an alternative path considering only the presence of real poles.  \\
\indent Therefore, considering asymptotic completeness in the whole Hilbert space, one would try to establish the physical states as states generated by a linear combination of a set of powers of the creation operator of the transverse part of the asymptotic massless field acting on the vacuum \cite{Nak1w}. Then, we provisionally assume the following subsidiary condition based on the asymptotic field 
\begin{equation} B^{+(0)}(x) | \text{phys} \rangle = 0,\quad  \big(\Box^2 A_\mu^{+\ as}(x)\big) | \text{phys} \rangle = 0 \quad \forall | \text{phys} \rangle \in \mathcal{V}_{\text{phys}}.
\label{3}
\end{equation}
present in the 
Yang-Feldman equation
\bea A_\mu(x)=A_\mu^{as}(x)+e\int d^4y P_{\mu \nu}^{(0) R/A}(x-y)\big(\bar \psi(y)\gamma^\nu \psi (y)\big)\eea with the $R/A$ labels denoting the retarded and advanced versions of the free-propagator obtained by  \eqref{prop} in the limit $\pi^R(p)\to 0$. Unfortunately, as we are going to see, the subsidiary condition \eqref{3} is not invariant under time evolution since in the interacting regime states containing massive ghosts can be created from a physical initial one. {The original idea was to consider the result from the interacting theory \eqref{prop} ensuring the existence of a massless pole even for the asymptotic renormalized structure. Then, taking into account the postulate of asymptotic completeness \cite{Nak1w}, all the states can be at least formally constructed upon applying a given series of normal products of asymptotic field operators on the vacuum, see the seminal works \cite{kallen1,kallen2} establishing the peturbative operator solution for QED$_4$. The last operator procedure was extended for higher derivative systems \cite{davi} providing a wider background for our approach. \\
\indent To properly clarify why the last approach is no longer valid for the interacting theory, we present a concise digression. It is worth mentioning that the Bopp-Podolsky model can be equivalently reformulated in a coupled second derivative structure based on two sets of massless and massive fields \cite{pod8w}. The quadratic piece can be partially diagonalized as \footnote{The kinetic part of the fermionic field as well as the auxiliary sector are left implicit, just the current operator is highlighted. $\theta_{\mu \nu}(\Box)$ denotes the transverse projector. }
\begin{align}
    {\mathcal{L}}=-\frac{1}{4}&F_{\mu \nu}(x)F^{\mu \nu}(x)-\frac{1}{2} \left[m^2C_\mu(x)C^\mu(x)+C^\mu(x)\Box \theta_{\mu \nu}C^\nu(x) \right]\nonumber \\ &+J_\mu(x)\big(A^\mu(x)+C^\mu(x)\big)\label{decoupled},
\end{align}
whereas both of massless $A_\mu(x)$ and massive field $C_\mu(x)$ interact with the current operator. Therefore, although the kinetic part can be diagonalized, the radiative corrections imply an infinite series of mixing terms between both sectors in the quantum action. It indicates that the interaction ties the massive and massless sectors and they cannot be described separately. Then, since the theory is Hermitian, the $S$-matrix is unitary in the whole Hilbert space, $SS^\dagger=\operatorname{I}$. However, the presence of the negative norm states does not allow a probabilistic interpretation. Namely, consider the transition probability between the states that we are provisionally defining as physical into a given physical eigenstate of the asymptotic Hamiltonian $P_{1 \to \sigma}=\langle \text{phys}\ 1|S|\sigma \ \text{phys}\rangle \langle \text{phys}\ \sigma |S^\dagger|1\ \text{phys}\rangle  $. One can prove that the sum over all possible transitions $\sum_\sigma P_{1 \to \sigma}=\sum_\sigma \langle \text{phys}\ 1|S|\sigma \ \text{phys}\rangle \langle phys\ \sigma |S^\dagger|1\ \text{phys}\rangle  $ is not equal to $1$ unless it can demonstrated that $\sum_\sigma |\sigma \ \text{phys}\rangle \langle \text{phys}\ \sigma |=\operatorname{I}$ in the time-evoluted physical subspace. Since the $S$-matrix is unitary, the positive norm of $| \ \text{phys}1\rangle$ is kept by the state $S| \ \text{phys}1\rangle$. However,  it can be expressed as a linear combination of positive and negative norm eigenstates of the asymptotic Hamiltonian $S| \ \text{phys}1\rangle=\sum_X\alpha_X|X\rangle+\sum_{X'}\beta_{X'}|X'\rangle $ with the first contribution regarding positive norm states that can also contain an even number of asymptotic massive ghost creation operator whereas the last is associated with the negative norm states. This feature is a consequence of the specific structure of this interaction implying that this candidate for a physical subspace is not invariant under temporal evolution.  For a normalized initial state, the unitarity of the $S$-matrix implies $1=\big(\sum_X|\alpha_X|^2-\sum_{X'}|\beta_{X'}|^2  \big)$. It means that a consistent probabilistic prescription cannot be established for theories containing negative norm states, unless the contributions associated with the ghosts always vanish. Namely, one cannot associate a probability to the squared modulus of the state coefficients. \\
\indent Interestingly, for the case of standard QED$_4$, the negative norm states related to the longitudinal pure gauge fields are not created by time evolution due to the transverse nature of the interaction. More specifically, the Ward-Takahashi identities show that the whole quantum corrections are orthogonal to the longitudinal sector. It ensures that a positive norm subspace can be defined in compliance with the dynamics and the probabilistic interpretation. Regarding higher derivative theories, Lee and Wick \cite{lw}, and also some contemporary approaches \cite{grav} consider the mass parameter $m$ above the particle production threshold in order to eliminate the massive ghost from the asymptotic states due to the induction of a complex mass pole, leading to a kind of unstable mode. However, this is a disputed point since the authors of the recent paper \cite{kugo} argue that, if one considers the correct definition of the delta function distribution for the case of imaginary arguments, the interaction can still produce this complex mass mode, leading to a relevant contribution to the scattering processes. In this manner, it is worth mentioning the alternative approach of \cite{confghost} based on the confinement of the massive ghost of a higher derivative gravity model in a generalized quartet-like mechanism based on the BRST structure associated with the space-time symmetries. \\
\indent Returning to our specific discussion, since defining subspaces that avoid the transverse ghost mode is not compatible with the QED$_4$-like interaction, one can consider the reference \cite{quad} introducing an interesting approach based on modifying the born rule to set a consistent probabilistic interpretation for the theory, including processes associated with the ghosts. Accordingly, consistency with time evolution requires the application of the Dirac-Pauli quantization scheme for the ghost fields. Moreover, the latter should be considered as a physical mode in such an approach. However, in systems like one-loop renormalized gravity, this mode is undesired and generally believed to be just a feature of the truncation order of the perturbative series. In this manner, we opt to keep the standard quantization variables and search for alternative possibilities. \\
\indent Regarding the Bopp-Podolsky model, since massive ghosts can be created by the $S$ matrix, one redefines the physical subspace just as

\begin{equation}
B^+(x) | \text{phys} \rangle = 0, \quad \forall | \text{phys} \rangle \in \mathcal{V}_{\text{phys}}. \label{longe} \end{equation} 
consistently eliminating states related to the pure gauge sector. It is worth highlighting that the full operator $B(x)$ is being taken into account to avoid the longitudinal modes of both massless and massive gauge fields.\\
\indent According to the discussion on the impossibility of avoiding the massive ghost, it is worth mentioning the case of the previously defined current operator $J_\mu(x)$ acting on the vacuum. The current operator solution \cite{davi} for the case of the Bopp-Podolsky theory is such that it can be entirely written in terms of an infinite series of asymptotic fields including the ones associated with the ghosts. Therefore, considering the subsidiary condition \eqref{3} and the algebra \eqref{comut1} and \eqref{comut2}, it implies that even the gauge invariant state $J_\mu(x)|\Omega\rangle$, with $|\Omega\rangle$ being the complete vacuum, also has contributions from states of negative norm. Truncating the operator solution for $J_\mu(x)$ in equation (56) of \cite{davi} until the first order in the fine structure constant, and taking into account the perturbative nature of the expansion is sufficient to conclude that the projection $\langle \Omega|J_\mu(x)\big(\bar{a}_\lambda(\vec p)\big)^\dagger|\Omega\rangle$, with $\big(\bar{a}_\lambda(\vec p)\big)^\dagger$ being the previously introduced creation operator of a transverse asymptotic massive ghost, is non-vanishing.\\ 
\indent Moreover, considering this last result, a spectral representation for $\langle \Omega|J_\mu(x)J_\nu(y)|\Omega\rangle$, up to the transverse tensor projector encoding the current conservation, presents an infinite set of negative-definite contributions. It can be understood by introducing the completeness relation $\operatorname{I}=\sum_X \int d\Pi_X|X\rangle \langle X|-\sum_{X'}\int d\Pi_{X'}|X'\rangle \langle X'|$, between the current operators present in the two-point function \cite{G6}, and taking into account that $\operatorname{I}$ is based on the asymptotic eigenstates of the Hamiltonian.
However, this last feature is not entirely problematic, since it is responsible for reducing the degree of divergence associated with this observable by consistently evading the Lehmann theorem \cite{Nak1w}.  

\subsection{Negative norm states and ultraviolet improvement}

\indent  At this point, it is instructive to discuss some aspects of the optical theorem, highlighting important aspects regarding the last considerations. It concerns constraints in the Hilbert space projections associated with the unitarity of the $S$-matrix. It unveils the importance of inserting the correct definition of the identity $\operatorname{I}$  in the formal computations of higher derivative generalized QED$_4$. Since $SS^\dagger=\operatorname{I}$, writing $S$ as $S=\operatorname{I}+i\mathcal{T}$, with $\mathcal{T}=e\int d^4x \bar \psi(x)\gamma_\mu \psi(x)A^\mu(x)$, implies, after including the identity and external state $|a\rangle$,  $\langle a|i\big(\mathcal{T}-\mathcal{T}^\dagger\big)|a\rangle=\sum_X \int d\Pi_X\langle a|\mathcal{T}|X\rangle \langle X|  \mathcal{T}^\dagger |a\rangle-\sum_{X'}\int d\Pi_{X'}\langle a|\mathcal{T}|X'\rangle \langle X'| \mathcal{T}^\dagger |a\rangle$.
Then, we consider the case of the one-loop electron self-energy in Bopp-Podolsky theory  \cite{pod8w}, since the contribution from the ghost occurs already at one loop level \footnote{We are just highlighting the non-longitudinal sector of the boson propagator because it is the only one that leads to non-trivial contributions for the associated amplitudes.}
\begin{equation}
     i\Sigma(p)=-e^2\int \frac{d^4k}{(2\pi)^4}\frac{\gamma^\mu\big(\slashed{k}+m_e\big)\gamma_\mu}{\big( k^2-m^2_e+i\epsilon \big)}\Bigg[\frac{1}{\big( (p-k)^2+i\epsilon \big)}-\frac{1}{\big( (p-k)^2-m^2+i\epsilon \big)}\Bigg]
    \end{equation}
    with the massive ghost propagator appearing with its characteristic negative residue.\\
\indent After attaching external electron states, one can use the Cutkosky rules to obtain the discontinuity of the amplitude \footnote{The label $\upalpha$ denotes the sum over external massive/massless photon polarizations and $\upalpha'$ is related to the sum over fermionic degrees of freedom. }    ${\bar{u}}(p)2\operatorname{Im}\big[\Sigma(p)  \big ] u(p)$ 
\begin{multline}
     \!\!\!\!\!\!\!\!\! -e^2\int \frac{d^4k}{(2\pi)^2}\theta(k_0)\theta(p_0-k_0)
\bar{u}(p) \gamma^\mu \big(\slashed{k}+m_e\big)\gamma_\mu u(p)\Big(\delta\big((p-k)^2\big)-\delta\big((p-k)^2-m^2\big)\Big) \delta \big(k^2-m^2_e\big) \Big)  \end{multline}
and, considering the polarization sum rules \eqref{pol1} and \eqref{pol2}, conclude that it is equal to 
\begin{align}
     &\sum_{\upalpha, \upalpha'}\!\int\!\frac{d^4k}{(2\pi)^2} \theta(k_0)\theta(p_0-k_0) \delta\left(k^2\!-\!m^2_e \right)\delta\left((p\!-\!k)^2\! \right)
       \vcenter{\hbox{\includegraphics[width=4.2cm,height=2.18cm]{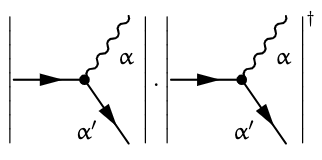}}}\nonumber \\
       - &\sum_{\upalpha, \upalpha'}\!\int\!\frac{d^4k}{(2\pi)^2} \theta(k_0)\theta(p_0-k_0) \delta\left(k^2\!-\!m^2_e \right)\delta\left((p\!-\!k)^2\!-\!m^2 \right)
       \vcenter{\hbox{\includegraphics[width=4.2cm,height=2.18cm]{decay-self.png}}}
       \end{align}
verifying the optical theorem. The minus sign in the last term is due to the correct insertion of the identity and is associated with the sum over the types of external massive photons in the completeness relation. As mentioned, the presence of the negative-definite contributions is related to an improvement of the renormalization properties of the $\Sigma(p)$ as revealed by its spectral representation. We are explicit in the present considerations to highlight important aspects in connection with the previous discussion on the UV improvement of observables such as the current-current two-point function. The latter is one of the most attractive properties of higher derivative field theories. \\
\indent Although the higher derivative structure implies UV enhancement, the presence of ghosts avoids the probabilistic interpretation. However, if the ghosts cannot become external asymptotic states, the previously mentioned probability concerning scattering processes between a physical ghost-free state $1$ into physical asymptotic eigenstates of the Hamiltonian $|\langle \text{phys} 1|S^\dagger| \sigma\  \text{phys} \rangle|^2$  is such that the sum over all transitions equals one, since $\sum_\sigma |\sigma \ \text{phys}  \rangle \langle \text{phys}\ \sigma|$, would become the completeness relation between physical states in a time-invariant manner. Then, the c-number contributions for the $S$-matrix are UV-enhanced due to the virtual ghost contributions while keeping the probabilistic interpretation. Considering an unstable ghost with $m>2m_e$ seems to possibly ensure this scenario, avoiding problematic states from the asymptotic spectrum. On the other hand, the authors of \cite{kugo} argue that due to the particle's complex frequencies, a regularizing function must be added to guarantee convergence at $t \to \pm \infty$. It implies that in loop computations one must use a generalized complex distribution instead of the delta function. One consequence is that this complex ghost mode can indeed be produced in scatterings, violating the probabilistic interpretation of the theory. One can also claim that this mode decays in other states and can not be measured in scattering processes. The point is that since $S$ is unitary in the whole Hilbert space, the negative norm of a state is invariant under time evolution. Therefore, if it decays, more ghost particles should be created to preserve the negative norm, defining the so-called anti-instability of such modes \cite{anti}. The positive message is that there exists a cutoff energy 
below which the ghost cannot be produced, and a physical unitary theory can be established. It reads $\Lambda=\operatorname{Re}\sqrt{M^2}-\operatorname{Im}\sqrt{M^2}$, with $M$ denoting the complex pole for $m>2m_e$. Then, the theory can consistently define an effective model until this energy limit.\\
\indent Finally, returning to the central issue of our initial investigation, we verified that the transverse ghost content cannot be excluded from a given subspace in compliance with the present interaction. Then, one may wonder if there are interactions compatible with a positive-definite subspace even in the presence of ghosts. We address it in the next subsection.

\subsection{An interacting model in compliance with a positive norm subspace}

\indent We have seen that for an interacting theory in the presence of transverse ghosts, one generally cannot set a positive norm space with a time-invariant definition. However, in this section, we present a model that complies with such procedure
\begin{align}
    {\mathcal{L}}=-&\frac{1}{4}F_{\mu \nu}(x)F^{\mu \nu}(x)-\frac{1}{2} \left[m^2C_\mu(x)C^\mu(x)+C^\mu(x)\Box \theta_{\mu \nu}C^\nu(x) \right]\nonumber \\ &+e\bar \psi(x)\gamma_\mu \psi(x)A^\mu(x)+g\big(\bar \psi(x)\sigma^{\mu \nu}\psi(x)F_{\mu \nu}^C\big)^2+B\p_\mu A^\mu+\tilde B(x)\p_\mu C^\mu(x)\label{decoupled},
\end{align}}
with $F_{\mu \nu}(x)$ being the field strength tensor for the $A_\mu(x)$ field and $F_{\mu \nu}^C(x)$ for the massive one. Here,  $\sigma_{\mu \nu}=\frac{i}{4}\Big[\gamma_\mu,\gamma_\nu\Big]$. The kinetic part of the fermion Lagrangian is left implicit. We also consider $m<2m_e$.\\
\indent In the absence of interaction, the system is equivalent to the free Bopp-Podolsky model. When one includes fermionic matter, a standard coupling is assumed for the massless field whereas the square of a Pauli-like coupling is included for the ghost. The latter is associated with a dimensionful coupling constant $g$ indicating a non-renormalizable interaction. Due to the $U(1)$ invariance of the system, the auxiliary fields are free $\Box B(x)=\Box \tilde B(x)=0$. Moreover, the model is invariant under the additional discrete symmetry 
\bea \mathcal{O}C_\mu(x)=-C_\mu(x)\ , \ \mathcal{O}\tilde B(x)=-\tilde B(x) \eea
which is also present in the massive ghost operator equation of motion
\bea m^2C^\omega(x)+\Box \theta^{\omega \nu} C_\nu(x)  =-g\p_\gamma [(\bar \psi(x)\sigma^{\mu \nu}\psi(x)F_{\mu \nu}^C)(\bar \psi(x)\sigma^{\gamma  \omega}\psi(x))]-\partial^\omega \tilde B(x)         \label{opsym}                        \eea
as it should be. One can directly verify  that the discrete symmetry represented by $\mathcal{O}$ can be consistently implemented in an iterative perturbative solution based on the asymptotic field operators.\\
\indent It is worth mentioning that the Fock eigenstates of the asymptotic Hamiltonian define a basis in the whole Hilbert space. Since one can partition it into even plus odd components under the operation above, with both subsets displaying the required properties to define a subspace, a candidate for the physical subsidiary condition can be discussed. Moreover, the $S$ matrix,  based on the Hamiltonian operator, also exhibits the same invariance.\\
\indent Then, it implies the rule $\langle \text{odd}|S|\text{even}\rangle=0$ for scattering between even and odd asymptotic states under the discrete symmetry operation, since $\big [S, \mathcal{O} \big]=0$. Therefore, no odd state is created from an initial even one, and vice-versa. \\
\indent In this manner, a time-independent positive\footnote{The positivity can be inferred by the fact that the Fock states define a basis with the even subspace being clearly associated with a superposition of states of positive norm. } norm subspace can be defined as  
\begin{equation} B^{+}(x) | \text{phys} \rangle = 0,\quad \tilde B^{+}(x) | \text{phys} \rangle = 0,\quad  \mathcal{O}| \text{phys} \rangle =| \text{phys} \rangle \quad \forall | \text{phys} \rangle \in \mathcal{V}_{\text{phys}}.
\label{3}
\end{equation}
with the zero norm states being consistently eliminated in the standard Cauchy completion of the quotient space definition \begin{equation}
\mathcal{H}_{\text{phys}}=\frac{\mathcal{V}_{\text{phys}}}{\mathcal{V}_0} \end{equation}
\indent This definition avoids pure gauge states and states that are odd under the operation $\mathcal{O}$. Owing to asymptotic completeness in the whole Hilbert space, all states can be constructed through the application of a series of powers of the asymptotic fields on the vacuum, implying that just the ones based on a superposition of states presenting an even number of the transverse part of $C_\mu^{\ as}(x)$ are within $\mathcal{V}_{\text{phys}}$. Moreover, considering just the standard QFT principles and the Hermitian nature of the interaction, one verifies that the physical $S$-matrix is also unitary. Since odd and even states are orthogonal, and $S$ does not create odd states from an initial even one, the sum over external asymptotic physical Hamiltonian eigenstates contained in the completeness relation $\sum_\sigma |\sigma\ \text{phys}\  \rangle \langle  \text{phys}\ \sigma| $  equals the identity in the physical subspace enabling a probabilistic interpretation according to the previous discussions.\\
\indent Therefore, although it is not possible to define a time-invariant ghost-free physical subspace, a positive definite one indeed is, due to the specific form of the interaction. It is also worth mentioning that the asymptotic Hamiltonian has positive definite matrix elements in this subspace. On the other hand, states with an even number of asymptotic ghost modes are within this subspace and can be created from a ghost-free one. They contribute to non-trivial processes as well as the other physical states.\\

\section{Conclusions and perspectives} \label{9}

\indent Throughout this article, the covariant operator quantization of two higher derivative systems was performed. The first is associated with a vector dual description of a spinless model. Issues on duality, reducible structure of the local freedom, and its massless limit were investigated. The analysis was restricted to $D=2+1$ dimensions in order to incorporate just two auxiliary fields. The (KON) quantization framework was successfully extended to Ostrogadskian systems. The choice of the subsidiary condition was suitable for eliminating all pure gauge states from the physical Hilbert subspace. It was also explicitly verified that the scalar massive pole degree of freedom indeed fulfills the required conditions to define an observable. Moreover, the higher derivative structure ensures a smooth massless limit.\\
\indent Regarding the generalized free higher derivative electrodynamics, the same processes were carried out. In this case, just one auxiliary field was necessary to quantize the system. The full set of commutators was obtained considering the correspondence principle and the operator equations of motion. A careful discussion on a suitable new kind of subsidiary condition to define the physical subspace was provided. This was necessary since, beyond the spurious gauge modes, the model also presents negative norm excitations in the transverse sector. Therefore, a set of two compatible conditions were considered to define the positive Hilbert subspace. The next achievement was related to the explicit verification of Hamiltonian positivity. Namely, although not positive-definite in the whole Hilbert space, it is indeed positive within the physical subspace.  \\
\indent Moreover, discussions on the interacting phase were also performed. A tentative definition of the physical subspace was provided considering asymptotic completeness. However, due to the nature of this specific interaction, massive ghost particles can be created from a ghost-free initial state. Therefore, defining a time-invariant subspace free of negative norm states is impossible in this situation. Accordingly, we provided a set of formal discussions on Lee-Wick correlated approaches based on recent contemporary discussions on higher derivative models. We also develop an alternative discussion based on the lack of full positivity of the contributions concerning a class of spectral densities, and its relation to the UV improvement characteristic of these systems.\\
\indent As a synthesizing achievement, defining one of the main objectives of the paper, the last subsection was devoted to introducing an interacting model that indeed complies with establishing a positive definite subspace even in the presence of ghost excitations. It was related to the existence of an additional discrete symmetry. The perturbative analysis of such a model can lead to additional insights in this discussion, defining a legitimate goal for a future investigation.  \\
\indent Therefore, as a future perspective, we proposed in suitable points throughout the article the consideration of part of the covariant operator approach based on the full Ostrogasdkian structure developed here to furnish an alternative way to address the problem of negative norm ghosts appearing in other higher derivative systems such as the one-loop renormalized linearized quantum gravity \cite{donogcosm}. Also, formal discussions like \cite{confghost} can furnish a base to address these issues considering a generalized quartet mechanism associated with the BRST-like quantum version of the space-time symmetries.\\
\indent Another possibility is investigating a higher derivative vector or tensor models presenting a Proca or Fierz-Pauli mass terms in addition to a Bopp-Podolsky-like term. For an interacting theory, the contributions from the self-energy can enrich this discussion by providing a set of phases depending on the ratio between mass parameters and the existence of particle production thresholds. Then, one can ask if a  Lee-Wick like \cite{lw} approach with the induction of complex mass poles through the interactions can be established in such a situation. Moreover, the investigation of the radiatively corrected interparticle potential between static charges or even energy-momentum sources can provide physical signatures of this mechanism related to these two classes of mass terms. 

\section{Appendix}
\indent Here, we analyze the Hamiltonian formulation of the dual vector spin-$0$ model in order to reinforce some conclusions achieved throughout the paper. We consider the Dirac-Bergman algorithm to derive the Hamiltonian positivity, the correct degrees of freedom, and a continuous massless limit. Let's consider the theory without the auxiliary fields 
\begin{align} {\mathcal{L}} =\frac{1}{2}\Big[\partial_\beta\big(\partial_\mu B^\mu\big)\partial^\beta\big(\p_\mu B^\mu\big)-m^2\big(\p_\mu B^\mu\big)^2     \Big]            \end{align}
\indent Considering the momentum definition \eqref{mom}, the Hamiltonian density obtained through the Legendre transform reads 

\begin{multline}
     \mathcal{H}=\frac{\big(\pi_0^B\big)^2}{2}-\pi_0^B\partial_i\dot{B}_i+p_i^B\dot{B}_i+p_0^B\dot{B}_0+\partial_j\big(\dot{B}_0+\partial_iB_i \big)\partial_j\big(\dot{B}_0+\partial_iB_i \big)+\frac{m^2}{2}\big( \dot{B}_0+\partial_iB_i  \big)^2      \end{multline}
with the primary constraint $\pi_i^B\approx 0$. The Latin indices denote spatial coordinates. All sums present in the Hamiltonian density are Euclidian ones.\\
\indent The Dirac-Bergman consistency algorithm leads to the following extra set of constraints
\bea -p_i^B+\partial_i \pi_0^B \approx 0 \ ; \ p_0^B\approx 0      \eea
\indent Therefore,  due to their first-class nature, they remove 14 degrees of freedom from the entire sixteen-dimensional Ostrogadskian phase space, leading to one degree of freedom in the configuration space characterizing a spin-$0$ particle. The quantity and the classification of the constraints remain the same at the massless limit. It implies the aforementioned continuous behavior at this limit.\\
\indent The Hamiltonian evaluated at the constraint surface is positive-definite
\bea
     \mathcal{H}=\frac{\big(\pi_0^B\big)^2}{2}+\partial_j\big(\dot{B}_0+\partial_iB_i \big)\partial_j\big(\dot{B}_0+\partial_iB_i \big)+\frac{m^2}{2}\big( \dot{B}_0+\partial_iB_i  \big)^2      \eea

\acknowledgements
G.B. de Gracia thanks FAPESP 2021/12126-5 grant for financial support and UFTM for the hospitality. A.A. Nogueira thanks UNIFAL for its hospitality in his temporary stay as visiting professor. Both authors thank prof. Dr. B.M. Pimentel (IFT-UNESP) for the lessons on the subtle aspects of QFT.\\ \\

\indent  Data Availability Statement: No Data associated in the manuscript.

 \end{document}